\documentclass[reprint,prd,superscriptaddress,showpacs,nofootinbib,floatfix,preprintnumbers]{revtex4-1}
\pdfoutput=1
\usepackage{graphicx}
\usepackage{epsfig}
\usepackage{amsmath,amssymb,hyperref,enumerate}
\usepackage{pstricks}
\usepackage{slashed}
\usepackage{amsfonts}
\usepackage{url}
\usepackage{color}
\usepackage{appendix}
\usepackage{bm}
\usepackage{subfig}
\usepackage{caption}
\hypersetup{colorlinks,citecolor= black,linkcolor= black}
\definecolor{nicered}{rgb}{0.7,0.1,0.1}
\definecolor{nicegreen}{rgb}{0.1,0.5,0.1}

\def\({\left(}
\def\){\right)}
\def\[{\left[}
\def\]{\right]}

\begin{document}
\preprint{INT-PUB-17-047, MI-TH-1773}
\title
{\Large \bf A Simple Testable Model of Baryon Number Violation:  Baryogenesis,  Dark Matter, Neutron-Antineutron Oscillation and Collider Signals}
\author
{Rouzbeh Allahverdi}
\affiliation
{Department of Physics and Astronomy, University of New Mexico, 
Albuquerque, NM 87131, USA}
\author{P. S. Bhupal Dev}
\affiliation{
Department of Physics and McDonnell Center for the Space Sciences, 
Washington University, St.~Louis, MO 63130, USA}
\author{Bhaskar Dutta}
\affiliation{
Department of Physics and Astronomy,
Mitchell Institute for Fundamental Physics and Astronomy,
Texas A\&M University,
College Station, TX 77843, USA
}
\date{\today}

\begin{abstract}
We study a simple TeV-scale model of baryon number violation which explains the observed proximity of the dark matter and baryon abundances. The model has constraints arising from both  low and high-energy processes, and in particular, predicts a sizable rate for the neutron-antineutron ($n-\bar{n}$) oscillation at low energy and the monojet signal at the LHC. We find an interesting  complementarity among the constraints arising from the observed baryon asymmetry, ratio of dark matter and baryon abundances, $n-\bar{n}$ oscillation lifetime  and the LHC monojet signal. There are regions in the  parameter space where  the $n-\bar{n}$ oscillation lifetime is found to be more constraining than the LHC constraints, which illustrates the importance of the next-generation  $n-\bar{n}$ oscillation experiments. 
\end{abstract}
\maketitle

\section{Introduction}\label{sec:intro}

Baryon number ($B$), an accidental global symmetry of the Standard Model (SM) Lagrangian, must be broken to dynamically generate the observed baryon asymmetry of the universe~\cite{Sakharov:1967dj}. At the level of effective theories for $B$ violation, the $\Delta B=1$ operators are typically of dimension-6 (or dimension-5 in supersymmetric models), so the amplitude scales as $\Lambda^{-2}$ (or $\Lambda^{-1}$), where $\Lambda$ is the associated new physics scale.  The stringent limits on the classic $\Delta B=1$ process of proton decay~\cite{Takhistov:2016eqm, TheSuper-Kamiokande:2017tit} imply that $\Lambda \gtrsim 10^{15}$ GeV, or the underlying new physics must be at the Grand Unified Theory (GUT) scale or above~\cite{Nath:2006ut}.  On the other hand, the $\Delta B=2$ operators are of dimension-9, so the amplitude scales as $\Lambda^{-5}$ and the associated new physics scale can be much lower, even in the few TeV range~\cite{Phillips:2014fgb}, while satisfying the experimental constraints on $\Delta B=2$ processes like di-nucleon decay~\cite{Takhistov:2016eqm, Litos:2014fxa} and neutron-antineutron ($n-\bar{n}$) oscillation~\cite{BaldoCeolin:1994jz, Abe:2011ky, Aharmim:2017jna}. This makes it plausible to envisage a TeV-scale model of $\Delta B=2$ that can be tested in laboratory experiments, while simultaneously solving the cosmological puzzle of baryogenesis, and possibly as a bonus, the apparent coincidence of baryon and dark matter (DM) abundances. 

Recently, a minimal TeV-scale extension of the SM has been proposed to this effect that explains the proximity of baryon and DM abundances~\cite{Allahverdi:2010rh, Allahverdi:2013mza}. This model introduces $B$-violating interactions via a set of ${\cal O}({\rm TeV})$ color-triplet scalars $X_\alpha$ 
and a singlet Majorana fermion $\psi$ that are coupled only to right-handed (RH) quarks. Baryogenesis occurs via out-of-equilibrium decays of the $X_\alpha$'s. The new fermion $\psi$ becomes stable, hence a viable DM candidate, when it has approximately the same mass as the proton.  Obtaining the correct DM relic abundance requires a non-thermal mechanism as the DM annihilation rate turns out to be much smaller than the thermal value of $\langle \sigma_{\rm ann} v \rangle = 3 \times 10^{-26}$ cm$^3$s$^{-1}$. The late decay of a modulus field can be the common source of non-thermal DM and baryogenesis~\cite{Allahverdi:2010im}. In this model,  the $B$-violating interactions do not generate any new contribution to the electric dipole moment of neutron or electron even at two-loop level.

Since the stability of DM in this model is tied to the that of proton, no additional symmetry is needed to ensure that the DM candidate is stable. As a result, the model naturally predicts monojet/monotop signals at the LHC and the parameter space of this model is currently being probed by the LHC data~\cite{CMS, thesis}. In addition, $B$ violation at the TeV scale results in potentially observable low-energy effects, most notably double proton decay and $n-\bar{n}$ oscillation.  

In this work, we perform a detailed study of this simple, predictive model to carve out the allowed parameter space that can be probed in the near future. We show that successful production of DM and baryon asymmetry via modulus decay sets interesting bounds on the modulus mass range and  the DM coupling to quarks.  We also show that the  parameter space of the model can be investigated at the LHC via monojet, dijet and a pair produced dijet resonance searches. The monojet signal in the allowed parameter space is mostly produced by a $b$ quark fusion with $d$ and $s$ quarks where the  $b$ quark partly arises from the gluon splitting which allows a $b$ jet to be associated with the monojet signal. 
There are regions in the  parameter space where we find the $n-\bar{n}$ oscillation lifetime to be most constraining. Thus, the LHC and  $n-\bar{n}$ oscillation provide a complementary search strategy to investigate this simple model of baryogenesis and DM at high- and low-energy frontiers, respectively. 

The plan of the paper is as follows: In section~\ref{sec:model}, we describe the model.  Section~\ref{sec:relic} explains the common origin of DM and baryon abundance in this model and the ensuing constraints on the parameter space. In Section~\ref{sec:nnb}, we discuss the model prediction for the $n-\bar{n}$ oscillation lifetime.  Section~\ref{sec:col} analyzes the LHC constraints from monojet, dijet and paired dijet searches, and their complementarity with the low-energy and cosmological constraints. Our conclusions are  given in  Section~\ref{sec:con}.

\section{The Model} \label{sec:model}

We start with the SM gauge group and add renormalizable terms with $B$ violation.  Gauge invariance then requires the introduction of new colored fields. A minimal setup includes two iso-singlet color-triplet scalars $X_{\alpha}$ (where $\alpha=1,2$) with hypercharge $+4/3$, which allows (potentially) $B$-violating interaction terms $X_\alpha^* d^cd^c$ in the Lagrangian (in the 2-component Weyl fermion notation), where $d$ (with hypercharge $-2/3$) stands for the RH down-type quark and $d^c$ (with hypercharge $+2/3$) for the corresponding anti-quark. At least two $X$ flavors are needed to produce a baryon asymmetry from the interference of tree- and loop-level decays of $X$ governed by the $X_\alpha^* d^cd^c$ term. However, this term alone is not sufficient to generate the asymmetry, because the net asymmetry vanishes after summing over all flavors of $d^c$ in the final and intermediate states~\cite{Kolb:1979qa}. We therefore add other renormalizable terms $X_\alpha \psi u^c$ involving a singlet Majorana fermion $\psi$ (with hypercharge 0), which plays the role of DM in this model and also has an important role in $n-\bar{n}$ oscillation, as discussed below. The interaction Lagrangian is given by~\cite{Allahverdi:2013mza, Dev:2015uca, Davoudiasl:2015jja}
\begin{eqnarray}
{\cal L}   & \ \supset \ & (\lambda_{\alpha i} X_\alpha \psi u_i^c +\lambda^\prime_{\alpha ij}X^*_\alpha d_i^c d_j^c + {m_\psi \over 2} \bar{\psi}^c \psi + {\rm H.c.}) \, \nonumber \\
& + & m^2_{X_\alpha} \vert X_\alpha \vert^2 + ({\rm kinetic ~ terms}) \, .
\label{lag1}
\end{eqnarray}
where $i, j$ are the quark flavor indices (the color indices are omitted for simplicity). Note that due to color anti-symmetry, 
only $\lambda^\prime_{\alpha ij}$ with $i\neq j$ are non-zero. Also note that without the presence of the Majorana mass term of $\psi$ in Eq.~\eqref{lag1}, one could have simply assigned $B=-2/3$ to $X_\alpha$ and $B=+1$ to $\psi$, so that the terms $X^*d^cd^c$ and $X\psi u^c$ (and obviously $X^*X$) do not violate baryon number. The Majorana mass term $\bar{\psi}^c \psi$ breaks $B$ by two units in this model.

For $m_\psi\ll m_{X_\alpha}$, which we assume to be the case for our subsequent discussion, one can integrate out $X_\alpha$ to write an effective 4-fermion interaction $\psi u_i^c d_j^c d_k^c$. This will induce the decays $\psi\to p+e^-+\bar{\nu}_e$ and $\psi\to \bar{p}+e^++\nu_e$, as long as $m_\psi > m_p+m_e$, where $m_p$ and $m_e$ are the masses of proton and electron, respectively.  On the other hand, the same 4-fermion interaction will induce the (rapid) proton decay $p\to \psi+e^++\nu_e$ if $m_p>m_\psi+m_e$, which is unacceptable. Therefore, the only viable scenario to make $\psi$ absolutely stable is when  $m_p - m_e \leq m_\psi \leq m_p + m_e$~\cite{Allahverdi:2013mza} (see also Ref.~\cite{McKeen:2015cuz} for some related discussion). The remarkable point is that the stability of  $\psi$ within this mass window is due to that of the proton, with no need to introduce any additional symmetry. This leads to an important property of the DM in this model,~i.e. $m_\psi\simeq m_p$. 

We would like to note that the spin-independent direct detection cross-section of $\psi$ is small, $\sigma_{\rm SI} \leq 10^{-16}-10^{-15}$ pb for $m_X\sim O$(TeV)~\cite{Allahverdi:2013mza}. This is due to the coupling of $\psi$ to a particular chirality of up-type quarks, which suppresses the spin-independent scattering cross section as $m_X^{-8}$. On the other hand, the spin-dependent cross-section is only suppressed by $m_X^{-4}$ and could be larger: $\sigma_{\rm SD} \leq 10^{-6}-10^{-5}$ pb for $m_X\sim O$(TeV), but still much below the current experimental limits~\cite{Akerib:2017kat}.

Similarly, the indirect detection signals from DM annihilation in this model will be very hard to disentangle from the overwhelming astrophysical background due to the low DM mass. Thus, as we will show below, the LHC experiments at the high-energy frontier and the $n-\bar{n}$ oscillation experiments at the low-energy frontier provide the only effective ways of testing this DM model. 


\section{Common Origin of Dark Matter and Baryon Abundance} \label{sec:relic}
 
For $m_\psi \simeq m_p$, the only kinematically available DM annihilation channel is $\psi\psi\to u^cu^c$ mediated by $t$-channel $X_\alpha$. The corresponding annihilation rate is given by 
\begin{align}
\langle \sigma_{\rm ann} v \rangle \ \sim \ \frac{\vert \lambda_{\alpha1} \vert^4}{8\pi} \frac{m^2_\psi}{ m^4_{X_\alpha}} \, . 
\end{align}
For $m_{X_\alpha} \sim {\cal O}({\rm TeV})$, even $\vert \lambda_{\alpha1} \vert \sim 1$ results in a rate that is much smaller than the thermal annihilation rate of $3 \times 10^{-26}$ cm$^3$s$^{-1}$ for obtaining the correct relic density. This implies that thermal freeze-out leads to overproduction of DM in this model, in accordance with the Lee-Weinberg bound~\cite{Lee:1977ua}. This requires a non-thermal production mechanism to explain the correct DM abundance.  

One possible scenario is the  late decay of a scalar field $\phi$ that reheats the universe to a sub-GeV temperature. A concrete example of this scalar field $\phi$ is the moduli, which typically arise in string compactification and supersymmetric models~\cite{Allahverdi:2010im}. They are displaced from the minimum of their potential in the early universe and start oscillating when the Hubble expansion rate is $H \sim m_\phi$ (where $m_\phi$ is mass of the modulus). Moduli are long lived due to their gravitationally suppressed couplings to other fields and dominate the energy density of the universe before decaying. Moduli decay rate is given by
\begin{equation} \label{decrate}
\Gamma_\phi \ = \ {c_\phi \over 2 \pi} {m^3_\phi \over M^2_{\rm Pl}} ,
\end{equation}
with $c_\phi \sim 0.01-1$ in typical string compactification scenarios, such as KKLT-type~\cite{Kachru:2003aw}, and $M_{\rm Pl}=2.4\times 10^{18}$ GeV being the reduced Planck mass. Moduli decay occurs when $\Gamma_\phi \sim H \simeq 0.33 g_*^{1/2}T^2/M_{\rm Pl}$. Thus, using Eq.~(\ref{decrate}), the reheat temperature of the universe after modulus decay is given by 
\begin{equation} \label{rehtemp}
T_{\rm R} \sim c_\phi^{1/2} \left({10.75 \over g_*}\right)^{1/4} \left({m_\phi \over 50 ~ {\rm TeV}}\right)^{3/2}  (3 \: {\rm MeV}) ,
\end{equation}
where $g_*$ denotes the number of relativistic degrees of freedom at $T_{\rm R}$. Preserving the success of Big Bang Nucleosynthesis (BBN) requires that $T_{\rm R} \gtrsim 3$ MeV. On the other hand, non-thermal DM production in our model sets an absolute upper bound of $T_{\rm R} < 1$ GeV. The corresponding range of the modulus mass from Eq.~(\ref{rehtemp}) is $m_\phi \sim (50-2500)$ TeV for $c_\phi\sim {\cal O}(1)$. 

The number density of DM particles (normalized by the entropy density $s$) produced by $\phi$ decay is given by 
\begin{eqnarray} \label{DM}
&& {n_\psi \over s} \ = \ {n_\phi \over s} {n_\psi \over n_\phi} \ = \ {3 T_{\rm R} \over m_\phi}\:  {\rm Br}_{\phi \rightarrow \psi} \ \equiv \  Y_\phi \: {\rm Br}_{\phi \rightarrow \psi}  \, ,  
\end{eqnarray}
where $Y_\phi\equiv n_\phi/s$ is the yield from moduli decay and ${\rm Br}_{\phi \rightarrow \psi}\equiv n_\psi/n_\phi$ denotes the number of $\psi$ quanta produced per $\phi$ quanta (either directly or as a secondary from decay of other particles, like $X_\alpha$, that are produced from $\phi$ decay). Successful non-thermal production of DM via moduli decay requires that this number density match the observed DM abundance in the universe
\begin{equation}
\left({n_\psi \over s}\right)_{\rm obs} \ \simeq \ 5 \times 10^{-10} \left({1 ~ {\rm GeV} \over m_\psi}\right).
\label{eq:npsi}
\end{equation}
For ${\cal O}({\rm MeV}) \lesssim T_{\rm R} \lesssim {\cal O}({\rm GeV})$ and $c_\phi \sim 0.01-1$, we find that $Y_\phi^{\rm min}\equiv 9 \times 10^{-9} \lesssim Y_\phi \lesssim 3\times 10^{-7}\equiv Y_\phi^{\rm max}$ in our model. Regarding ${\rm Br}_{\phi \rightarrow \psi}$, we typically expect $10^{-2} \lesssim {\rm Br}_{\phi \rightarrow \psi} \lesssim 1$. Here the lower bound corresponds to the case when $\phi$ democratically decays to all of the particles in the model (i.e., SM particles plus $X_\alpha$ and $\psi$), and the upper bound corresponds to the extreme case when $\phi$ dominantly decays to $\psi$ (directly or indirectly). We then see from Eq.~(\ref{DM}) that moduli decay can yield the correct DM abundance in non-thermal fashion in our model.  

Regarding baryogenesis, the interference of the tree- and one-loop level diagrams in out-of-equilibrium decay of $X_\alpha\to \psi u_i^c, d_i^cd_j^c$ can generate a baryon asymmetry, provided that the couplings $\lambda_{\alpha i}$ and/or $\lambda^{\prime}_{\alpha ij}$ have $CP$-violating phases and there are at least two $X_\alpha$ flavors (i.e. $\alpha=1,2$). Note that the $X_\alpha$ decays can directly generate the baryon asymmetry, independent of the electroweak sphalerons, and therefore, can be realized at temperatures either above or below the electroweak scale. In the latter case, this provides a concrete example of the post-sphaleron baryogenesis mechanism~\cite{Babu:2006xc,  Babu:2008rq, Babu:2013yca}.  In the non-thermal scenario discussed above, the number density of $X$ particles depends on the moduli decay rate; therefore, the final baryon asymmetry also depends on the number of $X$ quanta produced per decay of $\phi$, parameterized by ${\rm Br}_{\phi\to X_\alpha}$, apart from the size of the $CP$ asymmetry $\epsilon_\alpha$ in $X_\alpha$ decays. We find that  
\begin{align}
\eta_B \ \equiv \ \frac{n_B-n_{\bar{B}}}{s} \ \simeq \ Y_\phi \sum_\alpha {\rm Br}_{\phi\to X_\alpha}\epsilon_\alpha \, .
\label{BAU}
\end{align}
This should be compared with the observed baryon asymmetry $\eta_B^{\rm obs}\simeq 10^{-10}$~\cite{Ade:2015xua}. Taking ${\rm Br}_{\phi \rightarrow X_\alpha}$ between $10^{-2}$ (for democratic modulus decay to all degrees of freedom in the model) and 1 (for modulus predominantly decaying to $X_\alpha$), and the allowed range of $Y_\phi\sim 10^{-8}-10^{-7}$ [see below~Eq.~\eqref{eq:npsi}], we find from Eq.~\eqref{BAU} that a relatively large $CP$ asymmetry $\epsilon \sim 10^{-3}-10^{-1}$ is needed to generate the observed baryon asymmetry. 

Such large values of $CP$ asymmetry can only be realized by a resonant enhancement mechanism, similar in spirit to the resonant leptogenesis scenario~\cite{Pilaftsis:2003gt, Dev:2017wwc}, where  the self-energy graphs dominate the $CP$-asymmetry for quasi-degenerate $X_\alpha$'s,~i.e.~when the mass difference is of the same order as their decay width: 
\begin{align}
\Delta m_X \ \equiv \ |m_{X_1}-m_{X_2}| \ \sim \ \frac{\Gamma_{X}}{2} \ll m_{X_{1,2}}\, ,
\label{res}
\end{align} 
where $\Gamma_X$ is the average decay width of $X_1$ and $X_2$. In our model, see~\eqref{lag1}, the tree-level decay width of $X_\alpha$ is given by  
\begin{align}
\Gamma_{X_\alpha} \ = \ \frac{m_{X_\alpha}}{16\pi}\left( \sum_i|\lambda_{\alpha i}|^2+\sum_{ij}|\lambda'_{\alpha ij}|^2 \right) \, .
\label{gamma}
\end{align}
The flavored $CP$ asymmetry is then given by~\cite{Dev:2014laa, Dev:2015uca} 
\begin{align} 
\epsilon_\alpha  \ = \ & \frac{1}{8\pi} \frac{\sum_{ijk}{\rm Im}(\lambda^*_{\alpha k}\lambda_{\beta k}\lambda'^*_{\alpha ij}\lambda'_{\beta ij})}{\sum_i |\lambda_{\alpha i}|^2+\sum_{ij}|\lambda'_{\alpha ij}|^2} \nonumber \\
& \times \frac{(m_{X_\alpha}^2-m_{X_\beta}^2)m_{X_\alpha}m_{X_\beta}}{(m_{X_\alpha}^2-m_{X_\beta}^2)^2+m_{X_\alpha}^2\Gamma_{X_\beta}^2} \,
\label{eps}
\end{align}
with $\alpha,\beta=1,2$ and $\alpha\neq \beta$.  In the exact resonant limit~\eqref{res}, it can be seen from Eq.~\eqref{eps} that the regulator (second term on the RHS) goes as $m_X/\Gamma_X$, where we have defined $m_X$ as the average mass of $X_{1,2}$.  From Eq.~\eqref{gamma}, we know that $\Gamma_X \propto m_X$, so the $CP$ asymmetry in the resonance limit becomes {\it independent} of $m_X$. This in turn implies that the baryon asymmetry given by Eq.~\eqref{BAU} also becomes {\it independent} of $m_X$, as long as $m_\phi\gg m_X$, so that the moduli decay is not kinematically suppressed. In what follows, we work in the resonance limit~\eqref{res} to calculate the $CP$ asymmetry for given values of the couplings.

%

For simplicity, we make a few reasonable assumptions on the flavor structure of the new couplings $\lambda_{\alpha i}$ and $\lambda'_{\alpha ij}$, taking into account the experimental constraints. In particular, the exchange of $X_\alpha$ in combination with the Majorana mass of $\psi$ leads to $\Delta B = 2$ and $\Delta ΔS = 2$ process of double proton decay $pp \rightarrow K^+ K^+$ in this model~\cite{Dev:2015uca}. For $m_\psi \sim {\cal O}({\rm GeV})$ and $m_X\sim {\cal O}({\rm TeV})$, current experimental limits on di-nucleon decay from Super-Kamiokande~\cite {Takhistov:2016eqm, Litos:2014fxa} imply that $\vert \lambda_{\alpha 1} \lambda^{\prime}_{\alpha 12} \vert \leq  10^{-6}$. To satisfy this bound, we assume that $\vert \lambda^{\prime}_{\alpha 12} \vert$ is very small (and drop it from our subsequent discussion), while $\vert \lambda_{\alpha 1} \vert$ can be ${\cal O}(1)$. 

Another $\Delta B = \Delta S = 2$ process is $p K^- \rightarrow {\bar p} K^+$, which is related to the double proton decay $pp\to K^+K^+$ via crossing symmetry. This is relevant for reactions involving cosmic rays and its cross-section goes like $\vert \lambda_{\alpha 1} \lambda^{\prime}_{\alpha 1 2} \vert^4 m^2_\psi/{\sf s}^2$ at center-of-mass energies $\sqrt{\sf s} \gg 1$ TeV. The limit from double proton decay then implies a cross-section below $10^{-64}$ cm$^2$, which is in agreement with the bounds from cosmic rays experiments~\cite{Patrignani:2016xqp}.     

The model also predicts the $\Delta B = 2$ process of $n-{\bar n}$ oscillation for which the relevant couplings are $\lambda_{\alpha 1}$ and $\lambda^{\prime}_{\alpha 13}$. The ensuing constraints will be discussed in Section~\ref{sec:nnb}. The couplings $\lambda^{\prime}_{\alpha 13}$ and $\lambda^{\prime}_{\alpha 23}$ can also lead to $\Delta S = 2$ processes of $K^0_s-K^0_{\bar s}$ and $B^0_s-B^0_{\bar s}$ mixing. However, color conservation does not allow any tree-level contribution to the mixing~\cite{Allahverdi:2010im, colorsup}, and hence, experimental bounds on these processes are easily satisfied in all of the allowed parameter space shown below. 
A combination of $\lambda_{\alpha i}$ and $\lambda^{\prime}_{\alpha 13},\lambda^{\prime}_{\alpha 23}$ also leads to novel monojet/monotop signals (depending on the relative size of $\vert \lambda_{\alpha 1} \vert, \vert\lambda_{\alpha 2}\vert$ and $\vert \lambda_{\alpha 3} \vert $), as well as dijet events, at the LHC~\cite{Dutta:2014kia, Allahverdi:2015mha}. The corresponding constraints from the latest LHC data will be discussed in Section~\ref{sec:col}.


To derive the cosmological constraints on the parameter space of the model, we choose a common value for all $\vert \lambda_{\alpha i}\vert$, denoted by $|\lambda|$. We also take $\vert \lambda^{\prime}_{1ij} \vert = \vert \lambda^{\prime}_{2 ij} \vert \equiv |\lambda^{\prime}_{ij}|$, with $\lambda^{\prime}_{12}$ being negligible as mentioned before. Modifications due to non-universality of these couplings will be straightforward to follow. Note that the pair of couplings $(\lambda_{1i},\, \lambda_{2i})$ or $(\lambda'_{1ij},\, \lambda'_{2ij})$ must have a relative phase between them, otherwise the $CP$ asymmetry in Eq.~\eqref{eps} vanishes. It is worth mentioning that successful baryogenesis can occur despite $\vert \lambda^{\prime}_{\alpha 1 2}\vert$ being very small in order to satisfy the di-nucleon decay and related bounds. In fact, we can seen from Eq.~(\ref{eps}) that a sizable $CP$ asymmetry can be generated as long as at least one of the $\vert \lambda^{\prime}_{\alpha 13}\vert$ and $\vert\lambda^{\prime}_{\alpha 23} \vert$ is large.   



\begin{figure}[t!]
\includegraphics[width=0.45\textwidth]{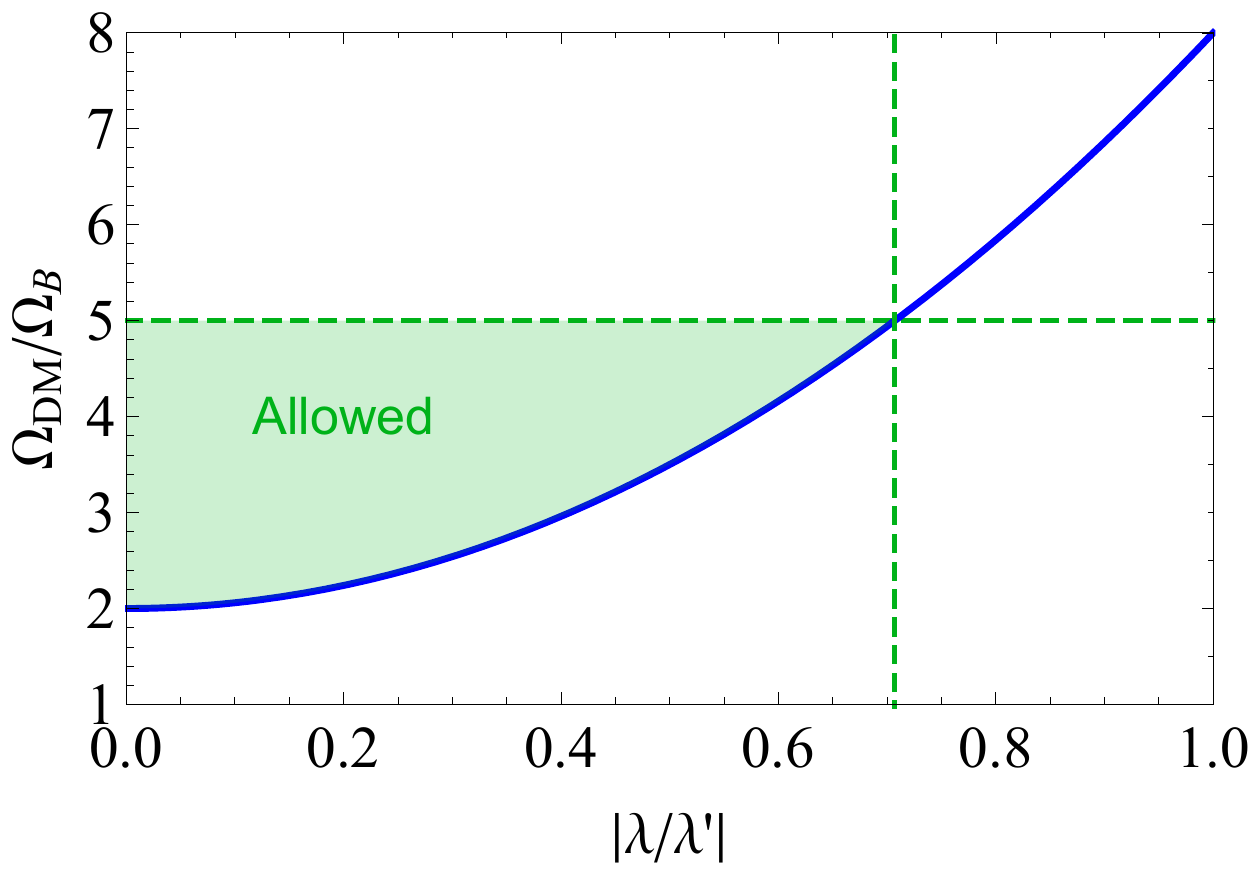}
\caption{Upper limit (vertical dashed line) on the ratio of couplings $|\lambda/\lambda'|$ from the DM-to-baryon ratio in the model (blue curve). The horizontal dashed line shows the observed value of this ratio. The green shaded region corresponds to the relation in Eq.~\eqref{coinc2}.} \label{fig:dmb}
\end{figure}

From Eqs.~(\ref{DM}) and (\ref{BAU}), we find the ratio of DM to baryon density (each normalized by the critical energy density of the universe):
\begin{align} \label{coinc}
{\Omega_{\rm DM} \over \Omega_{\rm B}} \ \simeq \ { {{\rm Br}_{\phi \rightarrow \psi} \over \sum_\alpha \epsilon_\alpha {\rm Br}_{\phi \rightarrow X_\alpha}}} \, .
\end{align}
We note that there are two contributions to ${\rm Br}_{\phi \rightarrow \psi}$ in Eq.~\eqref{coinc}: (i) due to the direct production of $\psi$ quanta from modulus decay (since the modulus has coupling to all fields), and (ii) due to the production of $X$ quanta from $\phi$ decay, followed by $X_\alpha$ decay to $\psi$. Since $m_\psi \ll m_X$, decay of $X_\alpha$ produces $\psi$ quanta with the following branching fraction:
\begin{align} \label{Br}
{\rm Br}_{X_\alpha \rightarrow \psi} & \ = \ {\sum_{i}{\vert \lambda_{\alpha i} \vert^2} \over \sum_{ij}{\vert \lambda^{\prime}_{\alpha ij} \vert^2} + \sum_{i}{\vert \lambda_{\alpha i}\vert^2}} \nonumber \\
& \ = \ {3 |\lambda|^2 \over |\lambda^{\prime }_{13}|^2 + |\lambda^{\prime }_{23}|^2 + 3 |\lambda|^2} \, .
\end{align} 
Thus, the numerator of Eq.~\eqref{coinc} can be written as 
\begin{align}
{\rm Br}_{\phi\to \psi} \ & = \ {\rm Br}_{\phi\to \psi}^{\rm direct}+\sum_\alpha {\rm Br}_{\phi\to X_\alpha}{\rm Br}_{X_\alpha\to \psi} \nonumber \\
\ & \geq \ \sum_\alpha {\rm Br}_{\phi\to X_\alpha}{\rm Br}_{X_\alpha\to \psi} \, ,
\label{Br2}
\end{align}
with $ {\rm Br}_{X_\alpha\to \psi}$ given by Eq.~\eqref{Br}, which is the same for $\alpha=1,2$ due to the simplified flavor structure considered above (the same is true for $\epsilon_\alpha$). From Eqs.~\eqref{coinc} and \eqref{Br2}, we find the following important inequality for the ratio of DM to baryon density in this model: 
\begin{align}
{\Omega_{\rm DM} \over \Omega_{\rm B}} \ \gtrsim \ \frac{{\rm Br}_{X_\alpha\to \psi}}{\epsilon_{\alpha}}
\label{coinc2}
\end{align}
%
Using Eqs.~\eqref{eps} and \eqref{Br}, the requirement that $\Omega_{\rm DM}/\Omega_{\rm B} \approx 5$~\cite{Ade:2015xua} sets an absolute {\it upper limit} on the ratio of couplings $|\lambda/\lambda'|\leq 1/\sqrt{2}$, where $|\lambda'|\equiv \sqrt{|\lambda'_{13}|^2+|\lambda'|_{23}^2}$. This is shown in Fig.~\ref{fig:dmb} by the intersection of the RHS of Eq.~\eqref{coinc2} (blue curve) and the observed value of $\Omega_{\rm DM}/\Omega_{\rm B}$ (horizontal line).  We emphasize that the upper limit on $|\lambda/\lambda'|$ is {\it independent} of the choice of $m_X, m_\phi$.  In Fig.~\ref{fig:dmb}, the green shaded region is allowed, because for values of $\Omega_{\rm DM}/\Omega_{\rm B}$ [as given by the RHS of Eq.~\eqref{coinc2}] less than the observed value, we could always arrange the matching contribution from the direct modulus decay [cf.~Eq.~\eqref{Br2}]. 

\begin{figure}[t!]
\includegraphics[width=0.45\textwidth]{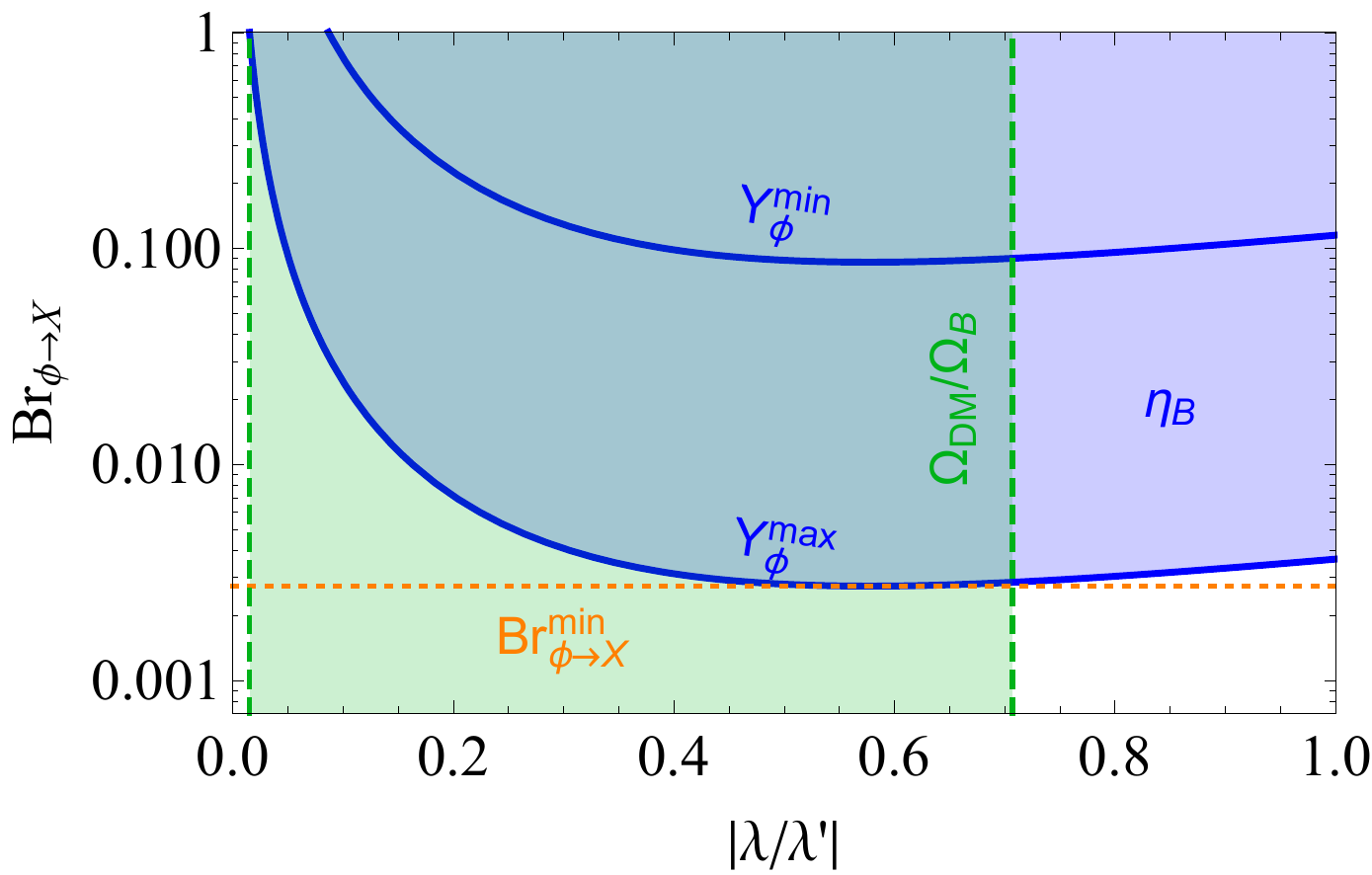}
\caption{Lower limit (left, vertical dotted line) on the ratio of couplings $|\lambda/\lambda'|$ from successful baryogenesis (blue shaded region). The horizontal dashed line shows the corresponding lower limit on the branching ratio ${\rm Br}_{\phi\to X_\alpha}$. The green shaded region is allowed by Eq.~\eqref{coinc2}.} \label{fig:baryo}
\end{figure}

It is clear from Eqs.~\eqref{BAU} and \eqref{eps} that obtaining the observed baryon asymmetry also imposes a {\it lower bound} on the same ratio $|\lambda/\lambda'|$. This is because the $CP$ asymmetry goes to zero in the limit of $|\lambda/\lambda'|\to 0$. This is shown in Fig.~\ref{fig:baryo} where various constraints are plotted in the ${\rm Br}_{\phi\to X_\alpha}-\vert\lambda/\lambda^{\prime}\vert$ plane. Here we superimpose the $\Omega_{\rm DM}/\Omega_{\rm B}$ constraint from Fig.~\ref{fig:dmb} (green shaded region) with the baryogenesis constraint (blue shaded region). The upper blue curve corresponds to the minimum allowed value of $Y_\phi$, whereas the lower blue curve corresponds to the maximum allowed value of $Y_\phi$ in the model. The region below the blue curves is excluded because it gives rise to a value of $\eta_B$ that is lower than observed value, even with the maximum possible value of $CP$ asymmetry (i.e., with resonant enhancement). In the region above the blue curves, one can always obtain the observed value of $\eta_B$ by departing from the resonant condition to lower the $CP$ asymmetry to the desired level. 

Two important results emerge from Fig.~\ref{fig:baryo}: (i) Since ${\rm Br}_{\phi\to X_\alpha}$ cannot be larger than 1, we obtain an absolute {\it lower limit} on $|\lambda/\lambda'|\gtrsim 0.02$, independent of $m_X$ and $m_\phi$. The lower bound becomes more stringent for smaller values of ${\rm Br}_{\phi\to X_\alpha}$.  (ii) The requirement of successful baryogenesis also imposes an absolute {\it lower bound} on the branching ratio of modulus decay to $X$ quanta: ${\rm Br}_{\phi\to X_\alpha}\gtrsim 2.7\times 10^{-3}$. This is independent of the details of any underlying string compactification model. 

\section{Neutron-antineutron oscillation} \label{sec:nnb}
\begin{figure*}[t!]
\includegraphics[width=0.45\textwidth]{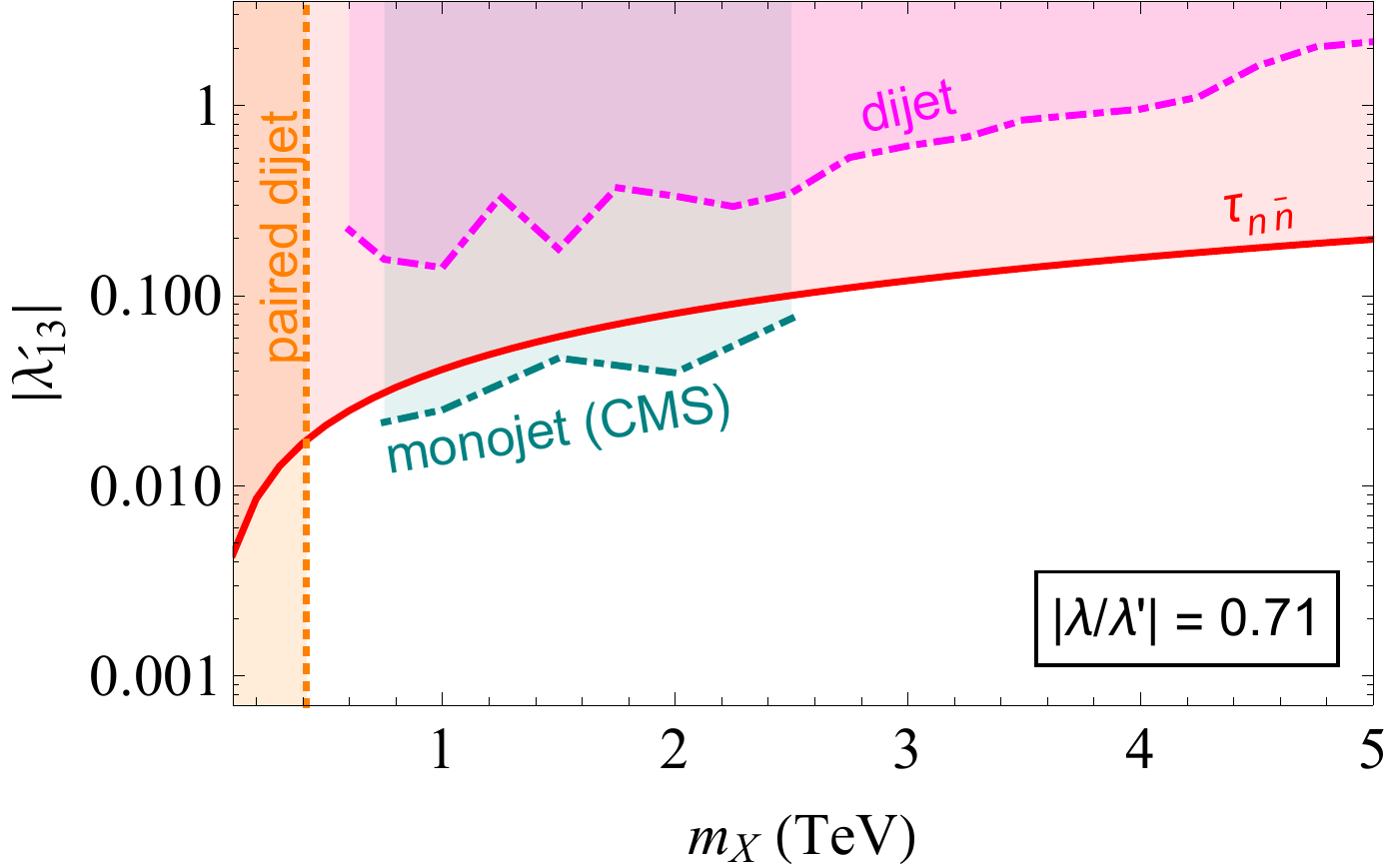}
\includegraphics[width=0.45\textwidth]{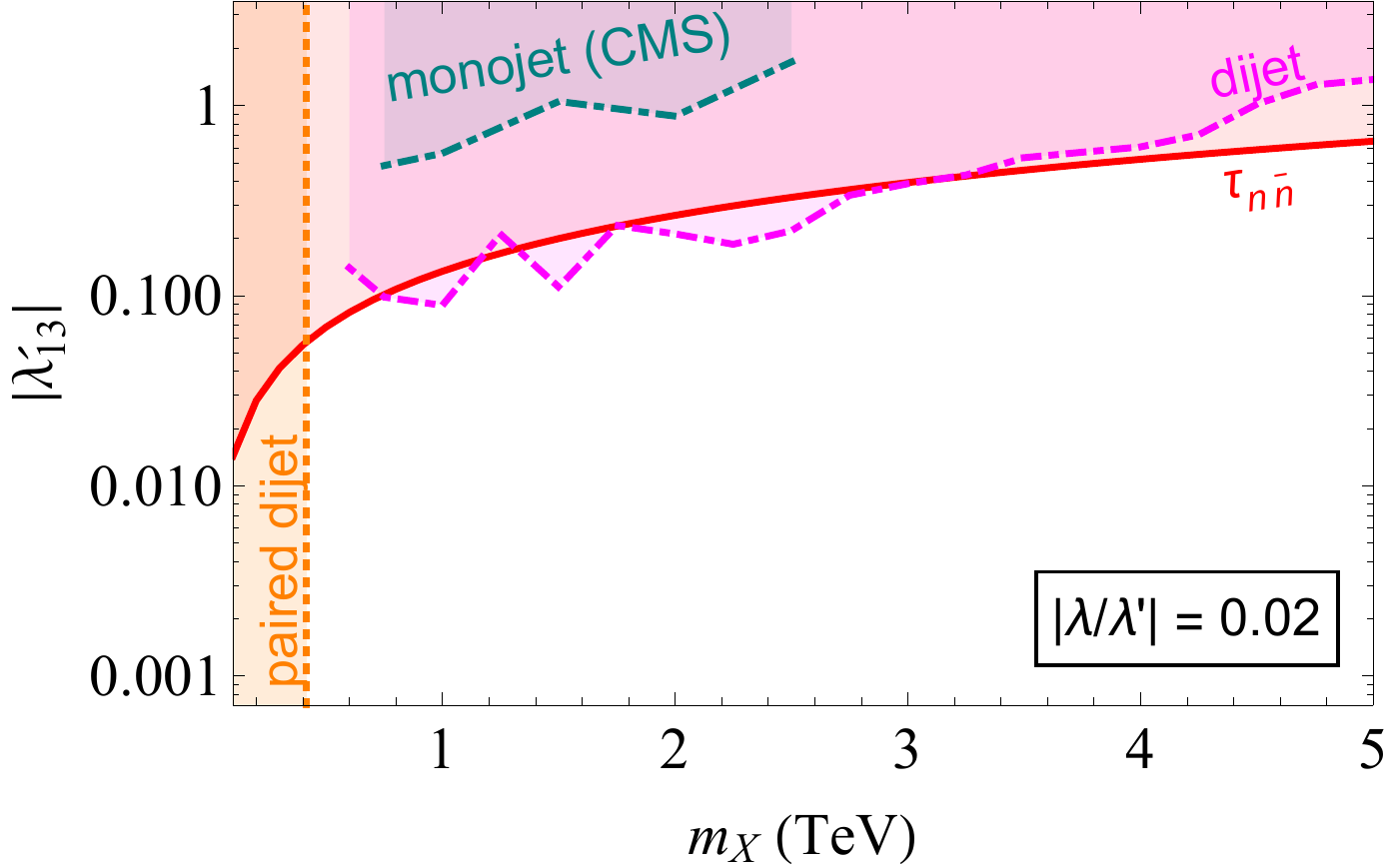}
\caption{The current $n-\bar{n}$ constraint (red, solid), compared with the monojet constraints from  
LHC (green, dot-dashed), dijet constraint (purple, dashed) and paired di-jet constraint (orange, dotted) derived using the $\sqrt {\sf s}=13$ TeV LHC data. The left and right panels are for $|\lambda/\lambda'|=0.71$ and 0.02, which are the maximum  and minimum values allowed by DM-to-baryon ratio (cf. Fig.~\ref{fig:dmb}) and baryogenesis (cf. Fig.~\ref{fig:baryo}) constraints, respectively. The shaded regions are ruled out at 95\% CL (except for the $n-\bar{n}$ constraint, which is at 90\% CL). }\label{fig:nnb-LHC}
\end{figure*}

After integrating out $X_\alpha$, the $B$-violating terms in Eq.~\eqref{lag1} induce an effective $B$-violating dimension-6 operator of the form~\cite{Babu:2006wz} 
\begin{align}
\frac{1}{m_{X_\alpha}^2} \lambda_{\alpha i}\lambda'_{\alpha jk}\psi u_i^c d_j^c d_k^c + {\rm h.c}.
\label{dim6}
\end{align}
Due to Majorana nature of $\psi$, the $\psi u^cd^cd^c$ term can give rise to $n-\bar{n}$ oscillation at the tree-level. However, because the couplings $\lambda'_{\alpha ij}$ are color antisymmetric, the two down-type quarks in~\eqref{dim6} must involve different families, which suppresses the tree-level contribution to $n-\bar{n}$ oscillation because of its dependence on the strange content of the neutron~\cite{Allahverdi:2010im}. The leading $\Delta B=2$ operator arises from the conversion of two strange or bottom quarks to two down quarks. This will need a $\Delta s=2$ or $\Delta b=2$ effective interaction. Due to the constraints on the $\Delta s=2$ operator from $pp\to KK$ lifetime~\cite{Takhistov:2016eqm, Litos:2014fxa}, the dominant contribution comes from the $\Delta b=2$ operator of dimension-9, which can be parameterized as
\begin{align}
\frac{m_\psi}{m_{X_\alpha}^6} \lambda_{\alpha 1}^2\lambda'^4_{\alpha 13}(u^cu^c)(d^cb^c)(d^c b^c).
\end{align}
This gives rise to $n-\bar{n}$ oscillation at one-loop level, 
with the amplitude given by~\cite{Dev:2015uca} 
\begin{eqnarray}
{G}_{n\bar{n}} \ \simeq \ \frac{\lambda^2 \lambda^{\prime 4}_{13} m_{\psi}}{16\pi^2 m_X^6}\ln\left(\frac{m_X^2}{m_\psi^2}\right) \, .
\end{eqnarray}
We translate this into the oscillation lifetime $\tau_{n\bar{n}}\sim (\Lambda_{\rm QCD}^6 {G}_{n\bar{n}})^{-1}$ and use the current experimental lower limit on $\tau_{n\bar{n}}\geq 3\times 10^{8}$ sec~\cite{BaldoCeolin:1994jz, Abe:2011ky, Aharmim:2017jna} to derive $n-\bar{n}$ constraints on the model parameter space. For concreteness, we fix the ratio $|\lambda/\lambda'|$ at the maximum and minimum allowed values of $1/\sqrt{2}$ and 0.02, respectively (cf. Fig.~\ref{fig:baryo}) to derive the constraints in the $m_X-\lambda'_{13}$ parameter space, assuming that $\lambda'_{23}=0$.\footnote{See Refs.~\cite{Babu:2006xc,  Babu:2008rq, Babu:2013yca, Dolgov:2006ay, Gu:2011ff, Calibbi:2017rab} for related studies connecting low-scale baryogenesis to $n-\bar{n}$ oscillation.} This already excludes a sizable portion of the parameter space, as shown by the red shaded regions in Fig.~\ref{fig:nnb-LHC}. The future sensitivity of  $\tau_{n\bar{n}}\geq 5\times 10^{10}$ sec, as predicted by next generation experiments~\cite{Phillips:2014fgb}, can probe even larger portion of the parameter space. 

\section{Collider Signals} \label{sec:col}
There is a novel monojet (or monotop, depending on the flavor structure of the $\lambda$ coupling) signal at the LHC from the on-shell production of $X$ and its subsequent decay to $\psi u^c$ through the $\lambda$-coupling: $pp\to X^{(*)}\to \psi u_i^c$, where the jet recoils against the DM particle and its transverse momentum has a Jacobian peak near one half of the resonance energy $\sqrt{\hat{\sf s}}=m_X$~\cite{Dutta:2014kia}. The cross section for this kind of monojet signal does not suffer from the high $p_T$ cut in typical LHC monojet searches for DM, where the jet comes from the initial state radiation. However, since we only have $\lambda'_{13}$ and $\lambda'_{23}$ couplings sizable in our model for the $Xd^c_i d^c_j$ type interactions, one of the initial states in the quark fusion process must include the $b$-quark, which has a relatively smaller parton distribution function (PDF) inside the proton, as compared to the light quarks. This PDF suppression is overcome by considering a gluon in the initial state (with much larger PDF), which splits into $b\bar{b}$ and one of the $b$'s then fuses with the $d$ quark in the other proton to produce $X$.\footnote{Recently, $b$-fusion has been used to search for $Z^\prime$ gauge boson mostly coupled to the third generation fermions~\cite{Dalchenko:2017shg}.} The additional $b$-quark in the final state can either be part of the inclusive monojet search (which just requires one high $p_T$ jet) or could be used to tag the signal events. In addition, the colored $X$ particles can be pair-produced via gluon-gluon fusion, each of which can subsequently decay into the DM and a $u^c$ quark. If one of the final state quarks satisfies the monojet selection criteria,  this would also contribute to the total monojet signal. To sum up, our monojet production cross-section has two separate contributions, namely, a single $X$ production from the $b-d$ or $b-s$ fusion, including the possibility of gluon splitting with an additional $b$ jet in the final state, and pair production of $X$ via gluon fusion. In the gluon fusion case, the additional $b$ jet does not exist.

A recent dedicated analysis from CMS~\cite{CMS, thesis} puts constraints on the $\lambda-\lambda'$ parameter space using the $\sqrt {\sf s}=13$ TeV monojet data with at least one jet with $p_T>100$ GeV, $|\eta|<2.5$, $E_T^{\rm {miss}}>250$ GeV and no leptons. In this analysis, the monojet production is assumed to have occurred via $X$ production due to $d-s$ quark fusion  and $gg$ fusion. However, as we have noted earlier, $\lambda'_{12}$ is small due to the double proton decay limit in our case and the monojet is produced from the $b-d$, $b-s$ and gluon fusions via single and pair production of $X$. In any case, the CMS limit from the dedicated analysis can be directly applied to our case. We first use the total cross-section limit as a function of $m_X$ as given in Ref.~\cite{thesis} which is shown by the green, dot-dashed  line in the $m_X-\lambda'_{13}$ plane in Fig.~\ref{fig:nnb-LHC}\ labeled as monojet (CMS). The CMS analysis  covers the mass range 750 GeV - 2.5 TeV. Here we have considered two different values of $|\lambda/\lambda'|=1/\sqrt{2}$ (left panel) and 0.02 (right panel), as for $n-{\bar n}$ oscillation discussed above. 
 It is interesting to see the complementarity between the low-energy $n-\bar{n}$ constraint (red solid line) and the high-energy monojet constraint. In particular, for smaller $X$ masses, the monojet constraint is more stringent, while for $m_X\gtrsim 3$ TeV, the production cross section at the LHC is kinematically suppressed, and the $n-\bar{n}$ process serves as a better probe in this part of the parameter space. For a smaller $|\lambda/\lambda'|$, the monojet constraint becomes significantly weaker, as the ${\rm Br}_{X\to \psi u^c}$ is suppressed compared to ${\rm Br}_{X\to d^c d^c}$. This is illustrated in the right panel of Fig.~\ref{fig:nnb-LHC}.

The ATLAS monojet analysis~\cite{Aaboud:2017phn} provides a constraint only on the effective theories of monojet production with upper limits on the cross-section for various missing energy cuts. Since this is not a dedicated analysis for the model we are using where the production cross-section depends on the mass scale of $m_X$, it is not possible to use ATLAS data in our case unlike what we have done with the CMS analysis. We expect, however, that if a similar analysis is done on the model using the ATLAS data, the constraint in the $m_X-\lambda'_{13}$ plane will be similar to the CMS constraint since both CMS and ATLAS have produced similar limits for monojet analysis on effective theories.


Apart from the monojet signal, the on-shell production of $X$ could also give a distinct dijet resonance through the $\lambda'$ coupling: $pp\to X^{(*)}\to d^c_i d^c_j$. We follow the recent CMS   analysis~\cite{Sirunyan:2016iap} (see also Ref.~\cite{Aaboud:2017wsi} for the corresponding ATLAS search) for dijet resonance search (above 600 GeV) with $\sqrt {\sf s}=13$ TeV LHC data and select the dijet events passing the cuts $p_T>30$ GeV, $|\eta|<2.5$, $m_{jj}>450$ GeV and $H_T>250$ GeV. Requiring our dijet signal cross section times acceptance to be compatible with the observed 95\% CL upper limit for the quark-quark case~\cite{Sirunyan:2016iap}, we derive the constraint shown by the purple, dashed line in Fig.~\ref{fig:nnb-LHC}, labeled as dijet. 
It is clear that for larger $|\lambda/\lambda'|$, the monojet constraint is better than the dijet one, while for smaller $|\lambda/\lambda'|$, the dijet constraint is better. This is simply due to the interplay between the corresponding branching ratios of $X$, namely, ${\rm Br}_{X\to \psi u^c}$ and ${\rm Br}_{X\to d^c d^c}$.

\begin{figure*}[t!]
\includegraphics[width=0.45\textwidth]{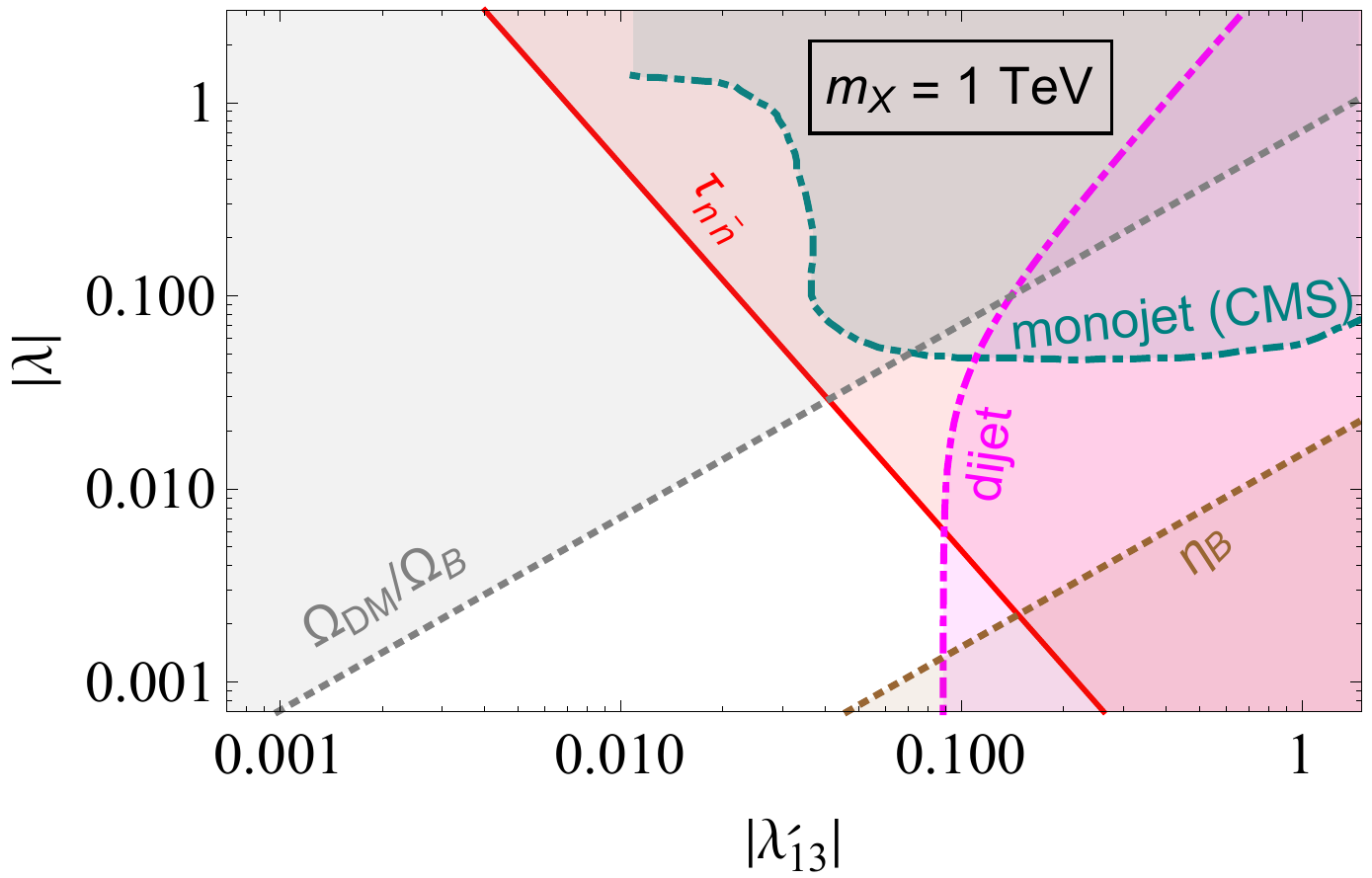}
\includegraphics[width=0.45\textwidth]{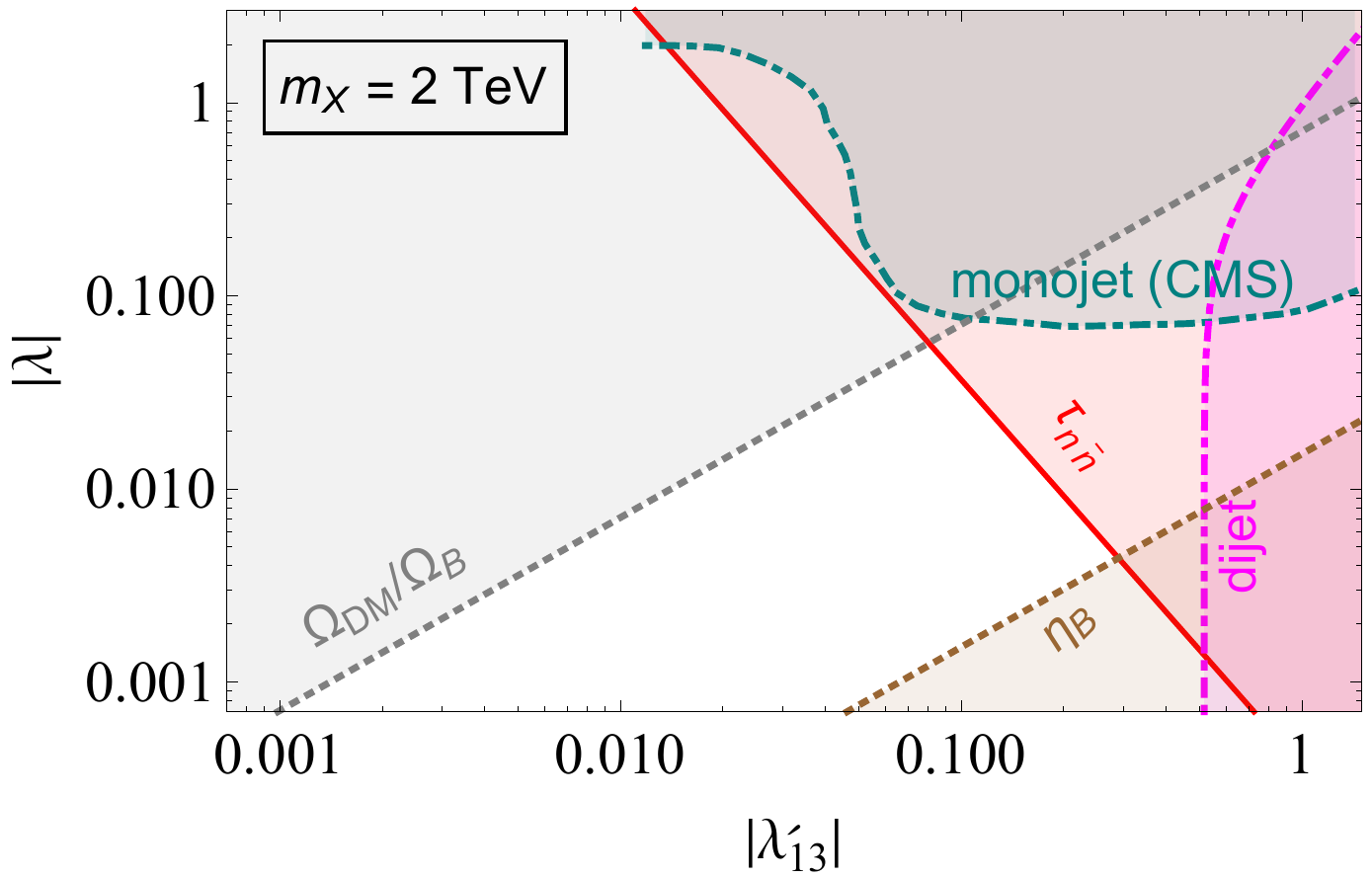}
\caption{The current constraints from $n-\bar{n}$ (red, solid), LHC monojet  (blue, dashed), dijet  (purple, dashed), $\Omega_{\rm DM}/\Omega_B$ (grey, dotted lines) and $\eta_B$(grey, dotted lines) are shown for $m_X=1$ TeV (left) and 2 TeV (right). The shaded areas are ruled out at 95\% CL (except for the $n-\bar{n}$ constraint, which is at 90\% CL). }\label{fig:nnb-LHC-cosmo}
\end{figure*}

The colored $X$ particles can also be pair-produced at a hadron collider: $pp\to XX^*$, purely through QCD interactions, independent of the $\lambda,\lambda'$ couplings. The subsequent decay of $X\to d^cd^c$ will then lead to a distinct paired dijet resonance. Following the recent ATLAS analysis for pair-produced resonances in four-jet final states~\cite{Aaboud:2017nmi}, we select events with at least four jets, each with $p_T>120$ GeV and $|\eta|<2.4$ and compare the signal cross section with the observed 95\% CL upper limit to obtain a lower limit on the mass of $X$. Note that this depends on $[{\rm Br}_{X\to d^c d^c}]^2$, which in turn only depends on the ratio $|\lambda/\lambda'|$ in our case. Therefore, for a given $|\lambda/\lambda'|$, the lower limit on the $X$ mass is independent of $\lambda'$, as shown by the vertical orange, dotted line in Fig.~\ref{fig:nnb-LHC}. This limit becomes stronger for a smaller $|\lambda/\lambda'|$. 

This model can also give rise to monotop final states with a RH top quark. This feature can be utilized to distinguish the model from the SM single top background which is consistent with left-handed top quarks~\cite{Allahverdi:2015mha}.

To further illustrate the complementarity between the high-energy constraints from LHC and low-energy constraints from $n-\bar{n}$ oscillation, along with the DM and baryogenesis requirements, we show in Fig.~\ref{fig:nnb-LHC-cosmo} the $\lambda-\lambda'$ parameter space for two benchmark values of the $X$ mass, $m_X=1$ and 2 TeV. As noted above, the DM-to-baryon-ratio requires $|\lambda/\lambda'|<1/\sqrt{2}$, whereas successful baryogenesis requires $|\lambda/\lambda'|>0.02$, independent of $m_X$, as shown by the  gray and brown dotted lines, respectively. The $n-\bar{n}$ constraint is shown by the red solid line, whereas the LHC constraint from monojet is shown by the green dashed line and from dijet  by the purple dot-dashed curves. The LHC constraints have been obtained with the recent $\sqrt {\sf s}=13$ TeV data following the procedure described above. As in Fig.~\ref{fig:nnb-LHC}, we find that for smaller $X$ masses, the LHC constraints are better, while for larger $X$ mass, the $n-\bar{n}$ constraint becomes more effective in probing the parameter space of this model.  




\section{Conclusion}\label{sec:con}
We analyzed a simple TeV-scale model of $B$-violation leading to successful baryogenesis, which also explains the apparent coincidence of the DM and baryon energy densities as due to a common origin from moduli decay. In this model, the DM particle is a Majorana fermion whose mass is required to be in a tiny window between $m_p-m_e$ and $m_p+m_e$, and its stability is linked to that of the proton with no ad-hoc discrete symmetry imposed. This naturally results in monojet/monotop signals at hadron colliders. Obtaining the observed DM-to-baryon ratio imposes an {\it upper} bound on the ratio of the couplings of the $B$-violating interactions involving the up-type and down-type quarks, while generating the observed baryon asymmetry imposes a {\it lower} bound on the same ratio, independent of the other model parameters. Neutron-antineutron oscillation is predicted at one-loop level, with a sizable rate that could probe a large and interesting part of the allowed parameter space. There exists a novel complementarity among the DM-to-baryon ratio, baryon asymmetry, $n-\bar{n}$ oscillation lifetime, and the LHC monojet and dijet signals in this model. 
The  existing  $n-\bar{n}$ oscillation lifetime constraint is already probing the parameter space for $m_X\gtrsim 2$ TeV, whereas the LHC monojet constraints are more relevant for the lower mass range. Thus, the ongoing LHC and  future $n-\bar{n}$ oscillation experiments provide a very important complementary search strategy to investigate this simple  model of baryogenesis and DM.

\section*{Acknowledgments}
The work of RA is supported in part by NSF Grant No. PHY-1417510. BD acknowledge support from DOE Grant de-sc0010813. PSBD would like to thank the organizers of the INT workshop on Neutron-antineutron oscillation, and especially, Kaladi Babu for stimulating discussions and comments on the preliminary results presented there. BD would like to thank Teruki Kamon and Shuichi Kunori for monojet-related discussions. RA and PSBD acknowledge the local hospitality provided by MITP, Mainz where this work was initiated.

\end{document}